# Large enhancement in hole velocity and mobility in p-type [110] and [111] silicon nanowires by cross section scaling: An atomistic analysis

Neophytos Neophytou and Hans Kosina

Institute for Microelectronics, TU Wien, Gußhausstraße 27-29/E360, A-1040 Wien, Austria

e-mail: {neophytou|kosina}@iue.tuwien.ac.at

## Abstract

The mobility of p-type nanowires (NWs) of diameters of $D$=12nm down to $D$=3nm, in [100], [110], and [111] transport orientations is calculated. An atomistic tight-binding model is used to calculate the NW electronic structure. Linearized Boltzmann transport theory is applied, including phonon and surface roughness scattering (SRS) mechanisms, for the mobility calculation. We find that large mobility enhancements (of the order of 4X) can be achieved as the diameter of the [110] and even more that of the [111] NWs scales down to $D$=3nm. This enhancement originates from the increase in the dispersion curvatures and consequently the hole velocities as the diameter is scaled. This benefit over compensates the mobility reduction caused by SRS as the diameter reduces. The mobility of the [100] NWs, on the other hand, is the lowest compared the other two NW orientations, and additionally suffers as the diameter scales. The bandstructure engineering techniques we describe are a generic feature of anisotropic bulk bands, and can be also applied to 2D thin-body-layers as well as other channel materials.

# I. Introduction

Silicon nanowire (NW) devices have recently attracted significant attention as candidates for a variety of applications such as high performance electronic devices [1, 2], efficient thermoelectric devices with potentially improved power factors and lower thermal conductivity [3, 4, 5, 6], optoelectronics [7, 8], and biosensors [9].

Low-dimensional materials offer the capability of improved performance due to the additional degrees of freedom in engineering their properties: i) the length scale of the cross section, ii) the transport orientation, iii) the orientation of the confining surfaces. In this work, we calculate the mobility of p-type Si NWs using the $sp^3d^5s^*$-spin-orbit-coupled (SO) atomistic tight-binding (TB) model and Boltzmann transport theory including phonon and surface roughness (SRS) scattering. The $sp^3d^5s^*$-SO model provides an accurate estimate of the electronic structure, while being computationally affordable. We examine cylindrical p-type NWs of diameters from $D$=12nm (approaching bulk) down to $D$=3nm, in [100], [110] and [111] transport orientations.

We find that cross section scaling of p-type Si NWs in the [110] and [111] directions largely increases their mobility by a factor of ~4X compared to bulk mobility. This is a consequence of a significant increase in the subbands' curvature and hole velocities with diameter scaling. SRS degrades the mobility of the NWs by ~2X for diameters below $D$=6nm. Still, however, the mobilities of the [110] and [111] NWs benefit from cross section scaling. The [100] NWs' mobility, is significantly lower than the [110] and [111] NW mobilities, and additionally it degrades with diameter scaling. Our results could potentially provide explanations for the large mobility enhancement in NWs that has been observed experimentally [10]. The bandstructure engineering techniques we describe are a generic feature of anisotropic bulk bands, and can be also applied to 2D thin-body-layers as well as other channel materials.

# II. Approach

The NWs' bandstructure is calculated using the sp$^3$d$^5$s*-SO TB model [11, 12, 13]. Each atom in the NW unit cell is described by 20 orbitals including spin-orbit-coupling. The model was extensively used in the calculation of the electronic properties of nanostructures with excellent agreement to experimental observations [14, 15, 16]. We consider here infinitely long, hydrogen passivated [17], cylindrical silicon NWs in three different transport orientations [100], [110], and [111] and diameters varying from *D*=12nm down to *D*=3nm. Relaxation of the NW surfaces and strain are neglected in this study.

The electronic structure of ultra-narrow NWs is sensitive to the diameter and orientation [13, 18, 19]. Figure 1 illustrates the dispersions of p-type NWs in the three different orientations [100], [110], and [111] (only half of the *k*-space is shown). The left column shows the dispersions of the *D*=3nm wires, and the right the dispersions of the *D*=12nm wires. The dispersions are all shifted to *E=0eV* for comparison purposes. Differences in the shapes, curvature and the number of subbands in the dispersions of the different wires can be observed. The origin of these features is explained in detail in Ref. [18], but the point here is that different electrical characteristics are expected for each wire due to the different electronic structures.

The linearized Boltzmann transport theory is used to extract the electronic properties for each wire using its dispersion relation. We consider electron-phonon and surface roughness (SR) scattering. The conductance is evaluated from:

$$\sigma = q_0^2 \int_{E_0}^{\infty} dE \left( -\frac{\partial f_0}{\partial E} \right) \Xi(E), \qquad (1)$$

where $\Xi(E)$ is defined as [20, 21]:

$$\Xi(E) = \sum_{k_x, n} v_n^2(k_x) \tau_n(k_x) \delta(E - E_n(k_x))$$
$$= \sum_n v_n^2(E) \tau_n(E) g_{1D}^n(E). \qquad (2)$$

Here

$$v_n(E) = \frac{1}{\hbar}\frac{\partial E_n}{\partial k_x} \tag{3}$$

is the bandstructure velocity, and $\tau_n(k_x)$ is the momentum relaxation time for a state in the specific $k_x$-point and subband $n$, and

$$g_{1D}^n(E_n) = \frac{1}{2\pi\hbar}\frac{1}{v_n(E)} \tag{4}$$

is the density of states for 1D subbands (per spin). The hole mobility ($\mu_p$) is defined as:

$$\mu_p = \frac{\sigma}{q_0 p}, \tag{5}$$

where $p$ is the hole concentration in the channel.

The transition and momentum relaxation scattering rates for a carrier in an initial state $k_x$ in subband $n$, to a final state $k_x'$ in subband $m$, is extracted from the electronic dispersions using Fermi's Golden rule [22]:

$$S_{n,m}(k_x, k_x') = \frac{2\pi}{\hbar}\left|H_{k_x',k_x}^{m,n}\right|^2 \delta\left(E_m(k_x') - E_n(k_x) - \Delta E\right), \tag{6}$$

$$\frac{1}{\tau_n(k_x)} = \sum_{m,k_x'} S_{n,m}(k_x, k_x')\left(1 - \frac{k_x'}{k_x}\right). \tag{7}$$

In this work we used the velocity $v_x(k_x)$ instead of the momentum $k_x$ in the last parenthesis of the equation above. The two are equivalent in the parabolic band case, but by using the velocity we can also capture curvature variation effects.

The total wavefunction of a specific state can be decomposed into a plane wave in the $x$-direction, and a bound state in the transverse plane $\vec{R}$. The total wavefunction can be expressed as:

$$\Psi_n(\vec{r}) = \frac{F_n(\vec{R})e^{ik_x x}}{\sqrt{\Omega}} \tag{8}$$

The matrix element between final and initial states then becomes:

$$H_{k_x',k_x}^{m,n} = \frac{1}{\Omega}\int_{u.c.} F_m^*(\vec{R})e^{-ik_x'x}U_S(\vec{R})F_n(\vec{R})e^{ik_x x}d^2R\,dx. \tag{9}$$

For the case of phonon scattering, the form factor is extracted by:

$$\left|I_{m,n}\right|^2 = \frac{1}{A}\int_R \left|F_m(\vec{R})\right|^2 \left|F_n(\vec{R})\right|^2 d^2R = \frac{1}{A_{mn}}. \quad (10)$$

In this work, for computational efficiency we consider the envelope function of the wavefunctions for the calculation of the form factor overlap. On each atom, we add the components of the squares of each multi-orbital wavefunction, and afterwards perform the final/initial state overlap multiplication. In such way, we approximate the form factor components by:

$$\sum_\alpha F_n^\alpha F_m^{\alpha*} \sum_\alpha F_n^\alpha F_m^{\alpha*} \approx \sum_\alpha F_n^\alpha F_n^{\alpha*} \sum_\alpha F_m^\alpha F_m^{\alpha*} = \frac{1}{A_{nm}^{\vec{R}}}, \quad (11)$$

where $\alpha$ runs over the TB orbitals on a specific atom. This treatment is equivalent to a single orbital (or effective mass) model, although we still keep the $k$-dependence. It reduces the memory needed in the computation by 20X, allowing simulations of large NW cross sections with only minimal expense in accuracy. In the cases of sine/cosine wavefunctions for square NWs, it can be shown analytically that for 1D intra-valley, the form factor $\frac{1}{A_{nm}} = \frac{9}{4A}$ [22, 23], where $A$ is the cross section area of the NW. For inter-valley overlaps $\frac{1}{A_{nm}} = \frac{1}{A}$. Indeed, these analytical expressions are good approximation of the numerical overlaps.

We include elastic acoustic phonons (ADP), inelastic optical phonons (ODP), and SRS. Confinement of phonons is neglected. Bulk phonons provide an ease of modeling and allow us to understand the effects on bandstructure on the mobility, still with good accuracy in the results. Spatial confinement mostly affects acoustic phonons. Its effect on the acoustic phonon limited mobility for the NW diameters examined in this work is of the order of 10-20% and declines fast as the diameter increases [24, 25, 26, 27]. These numbers, however, strongly depend on the boundary conditions one uses for the calculation of the phonon modes, which introduces an additional uncertainty [24, 26]. Optical phonons are not affected significantly by confinement [26, 28]. Clearly, the presence of additional scattering mechanisms (notably interface roughness scattering) also limit the mobility, and the relative effect of acoustic phonon confinement on the total mobility will be even smaller. Therefore, we believe that the bulk phonon approximation would not affect our results significantly.

For ADP scattering the rate is:

$$\frac{1}{\tau^n_{ADP}} = \frac{2\pi}{\hbar} \frac{D_{ADP}^2 k_B T}{\rho v_s^2} \sum_m \frac{1}{A_{nm}} g^m_{1D}(E), \tag{12}$$

where $g^m_{1D}(E)$ is the one dimensional density of final states per spin, $D_{ADP} = 5.39\text{eV}$ is the acoustic phonon deformation potential, $\rho$ is the mass density, and $v_s$ is the speed of sound in silicon. Although not specifically written in Eq. (12), or Eq. (13) and (15) further on, the *k*-dependence of the form factor as well as the velocity relaxation term shown in Eq. (7) are included in the calculation and should be assumed.

For ODP scattering, the rate is:

$$\frac{1}{\tau^n_{ODP}} = \frac{\pi}{\hbar} \frac{D_{ODP}^2 \left(N_\omega + \frac{1}{2} \mp \frac{1}{2}\right)}{\rho \hbar \omega_{OP}} \sum_m \frac{1}{A_{nm}} g^m_{1D}(E \pm \hbar \omega_{OP}), \tag{13}$$

where $D_{ODP} = 13.24 \times 10^{10} \text{eV}/\text{m}$ is the scattering amplitude, $\hbar \omega_{OP} = 0.62 eV$ is the phonon energy, and $N_\omega$ is the number of phonons given by the Bose-Einstein distribution. The parameters used are the same as in Ref. [24]. The $D_{ODP}$ used is higher than the bulk value $D_{ODP}^{bulk} = 6 \times 10^{10} \text{eV}/\text{m}$ [22]. Larger deformation potential values are commonly used to explain mobility trends in nanostructures [29, 30, 31]. Buin *et* al. [24] calculated mobilities of NWs up to *D*=3nm using the same model. In this work, by using the same parameters we were able to benchmark our results at least for the *D*=3nm case with a good agreement, before extending to larger diameters. The most accurate deformation potentials for NWs may finally be obtained by comparing mobility with experimental data as described in [24, 29, 31], which for NWs at this time are sparse. Using different sets of parameters will change the absolute values of our results, but not the mobility *trends* with diameter and orientation that we present. A comparison of mobilities extracted using different parameter sets is provided in the Appendix.

For SRS, we assume a 1D exponential autocorrelation function [32] for the roughness given by:

$$\langle \delta(\rho)\delta(\rho'-\rho)\rangle = \Delta_{rms}^2 e^{-\sqrt{2}|\rho|/L_C} \tag{14}$$

with $\Delta_{rms}$ = 0.48nm and $L_C$ = 1.3nm [30]. The scattering strength is derived from the shift in the band edges with quantization $\frac{\Delta E_V}{\Delta D}$ [33, 34]. Although this treatment ignores the effect of the wavefunction shape deformation on the interface, it is a valid approximation for ultra scaled channels, where the dominant SRS mechanism is the band edge variation [34, 35, 36]. The scattering rate is derived as [33, 35]:

$$\frac{1}{\tau_{SRS}^n} = \frac{2\pi}{\hbar}\left(\frac{2\Delta_{rms}^2 L_C}{1+\beta^2 L_C^2}\right)\left(\frac{q_0 \Delta E_V}{\Delta D}\right)^2 \sum_m g_{1D}^m(E) \tag{15}$$

where $\beta = k_x - k_x'$.

## III. Results and Discussion

Figure 2 shows the low field, phonon-limited mobility of NWs in the [100] (dotted), [110] and [111] (circled) transport orientations, as a function of the NW diameter from $D$=12nm down to $D$=3nm. The mobilities for larger diameters are below the silicon bulk hole mobility $\mu_p$=450cm$^2$/Vs, since the inter-valley optical phonon scattering amplitude used is larger than the one used to match the bulk mobility. Using the lower value will increase the large diameter mobilities and bring them closer to $\mu_p$=450cm$^2$/Vs. Nevertheless, our results demonstrate a clear increasing mobility trend as the diameter reduces. The increase is observed in the cases of [110] and mostly in the case of the [111] NWs, where the mobilities reach up to ~1300 cm$^2$/Vs (in agreement with the work of Ref. [24]) and ~2200 cm$^2$/Vs, respectively. These are a three-fold and five-fold increase from the bulk silicon mobility values, respectively. Comparing the $D$=3nm to the $D$=12nm values, the increase is about eight-fold. It is a demonstration of how the length scale degree of freedom can provide a mechanism of performance enhancement in nanostructures. On the other hand, the mobility of the [100] NWs is by far the lowest for all diameters, and additionally it decreases with diameter scaling.

The large increase in the hole mobilities of the [110] and [111] NWs is attributed to the enhancement of the carrier velocities, caused by increase of the subband curvature as the diameter reduces [19], (Fig. 1). Using a simple semiclassical ballistic model [13, 22, 37], we calculate the average bandstructure carrier velocity in these NWs as a function of the diameter, under non-degenerate conditions. The hole velocity increases by factor ~2X as the diameter reduces in the [110] and [111] NW cases as shown in Fig. 3. On the other hand, in the [100] case it remains almost constant, and therefore, no drastic changes in the mobility are observed either [18, 19].

The dispersions that determine the velocities in these channels are mixtures of subbands originating from the heavy-hole, the light-hole and the split-off bands and also include all band mixing, valley splitting, and quantization effects. These cannot be separated in a trivial manner. A single value of an approximate transport effective mass that will include the combination all the various bands, however, can be a useful quantity that would partly describe the electronic properties of the NW channels. The carrier velocities can be extracted from:

$$\left\langle v_x^+ \right\rangle = \frac{\sum_{k_x>0} v_x(k_x) f(k_x)}{\sum_{k_x>0} f(k_x)}. \quad (16)$$

After converting the sums into integrals and integrating over the positive $k_x$-space, the average carrier velocity can be evaluated as:

$$\left\langle v_x^+ \right\rangle = \sqrt{\frac{2k_B T}{\pi m^*}} \frac{\mathfrak{I}_{(d-1)/2}}{\mathfrak{I}_{(d-2)/2}}, \quad (17)$$

where $\mathfrak{I}_n$ is the Fermi-Dirac integral of order $n$, and $d$ is the dimensionality of the system ($d=1$ in this case). In the non-degenerate limit, an approximate transport effective mass can be extracted [22, 38]:

$$\left\langle m_{eff}^* \right\rangle = \frac{2k_B T}{\pi \left\langle v_x^+ \right\rangle^2}.  \qquad (18)$$

In our case we include several subbands of possibly different effective masses. The $\left\langle m_{eff}^* \right\rangle$ versus diameter is plotted in Fig. 4 for the three NW orientations considered. As expected, the values follow the inverse trend of the velocities shown in Fig. 3. The [111] and secondly the [110] channels have the lowest masses. Under strong scaling at $D$=3nm, they can be even lower than the electron transverse effective mass in the conduction band ($m^* = 0.19m_0$). This explains why the mobilities of these channels are much higher than that of bulk p-type channels (at least the phonon-limited mobility, since $\mu = q\tau / m^*$). In such case, the mobility benefits from: i) the mass decrease, and ii) an increase in $\tau$, as the lower density of states of the lighter subbands reduces the scattering rates.

Although the mobility of the [110] and [111] NWs largely increases as the diameter reduces, in reality, channels with thickness below 6nm suffer from enhanced SRS [34] that degrades the mobility. Figure 5 demonstrates the effect of SRS on the mobility. The solid lines are the low-field phonon-limited mobility results as in Fig. 2, whereas the dashed lines include both phonons and SRS. At larger diameters, the two curves coincide since SRS does not severely affect channels with thicknesses larger than 6nm. For the smallest diameters, however, SRS degrades the mobility by almost ~2X in all NW cases. The magnitude of this degradation is in qualitative agreement with other works on the mobility degradation of SRS in nanodevices [30], although it is subject to the parameters used in Eq. (15). Nevertheless, the [110] and [111] channels, that provide increasing carrier velocities as the channel width decreases, can partially compensate for SRS and still provide mobility benefits. As an example, we mention that an effective way to design high efficiency nanostructured thermoelectric devices is to scale the feature sizes in order to reduce the lattice thermal conductivity. In that scope, efficient thermoelectric devices based on silicon NWs have already been demonstrated [5, 6]. High electronic conductivity though is still needed. The [110] and [111] p-type channels, in which the mobility (and conductivity) increases with feature size reduction, might be suitable for such applications.

Although the enhancements in carrier velocity of [110] and [111] NWs originate from the NWs' cross section reduction, not all confining surfaces have the same properties [39, 40], nor provide the same benefits. To demonstrate the anisotropic effect of surface quantization in NWs, Fig. 6 shows the electronic structure of two *rectangular* [110] NWs, with [1$\bar{1}$0]- width and [001]-height directions, for two different cases. In the first case, the [1$\bar{1}$0]-width is *W*=3nm, whereas the [001]-height is *H*=12nm (Fig. 6a). In the second case the aspect ratio is reversed, namely *W*=12nm and *H*=3nm (Fig. 6b). The dashed lines indicate the first four subbands (two double-degenerate) of the *W*=*H*=12nm [110] NW. In the first case, strong confinement by the (1$\bar{1}$0) surface changes the dispersion such that the subbands acquire a large curvature, and therefore higher hole velocities. In the second case, (001) surface scaling does not provide this advantage and the subbands retain their small curvature and lower hole velocities as in the *W*=*H*=12nm NW. This behavior originates from the anisotropic shape of the bulk heavy-hole subband. The inset of Fig. 6a shows the heavy-hole (100) energy surface (with $k_z$=0) of bulk silicon. The 45º inclined lines indicate the relevant energy subbands/surfaces that will form the dispersion of NW or a thin-layer channel under (1$\bar{1}$0) surface quantization [13, 18]. The stronger the confinement, the farther away from the center the lines shift. Because of the anisotropic shape, the relevant bands acquire larger curvatures and provide higher hole velocities. The actual quantization involves significant band mixing, however, this analysis provides an indication as to where the main dispersion features originate from [18]. It is also worth mentioning that experimental data on silicon MOSFETs support the calculations for the advantageous p-type (1$\bar{1}$0)/[110] channel mobility over the (001)/[110] one [40].

Figure 7 shows the effect of the different surface confinement for the [111] NWs. In this case the relevant surface orientations are (1$\bar{1}$0) and (11$\bar{2}$). The dispersion of Fig. 7a is that of a *W*=3nm, *H*=12nm NW, in which the (11$\bar{2}$) surface dominates confinement. The dispersion of Fig. 7b is that of a *W*=3nm, *H*=12nm NW, in which the (1$\bar{1}$0) surface causes the strongest confinement. The dashed lines indicate the highest subband of the $W=H$=12nm [111] NW. Confinement of either surface is advantageous to the carrier velocities by increasing the subband curvatures (deviations from the dashed lines), with the $(11\bar{2})$ confinement being somehow more beneficial [19].

## V. Conclusion

The mobility of p-type silicon NWs in different transport orientations for diameters from $D$=12nm down to $D$=3nm is examined. The $sp^3d^5s^*$-SO atomistic TB model was used for electronic structure calculation. The linearized Boltzmann approach including phonon and SR scattering is used for transport calculations. The phonon-limited mobility in [110] and [111] NWs is shown to greatly increase by a least ~4X compared to the bulk value as the diameter scales down to $D$=3nm. This is a result of a large increase in the curvature of the dispersions and consequently increased hole velocities as the diameter scales. Although SRS degrades the mobility at diameters below $D$=6nm by ~2X, the large increase in the carrier velocities can compensate for this. Confinement can, therefore, still provide mobility enhancement. The [100] oriented NWs, on the other hand, have the lowest mobilities in all diameter ranges, and additionally, their mobility degrades with diameter scaling. Our results could potentially provide explanations for the large mobility enhancement in NWs that has been observed experimentally [10]. Finally, we mention that similar confinement caused velocity benefits can also be observed for 2D thin-body silicon layers and other channels materials with highly anisotropic bands.

## Acknowledgements
This work was supported by the Austrian Climate and Energy Fund, contract No. 825467.

# Appendix

The deformation potential values for nanostructures vary somewhat in the published literature. Here we compute the phonon limited mobility versus diameter for the [111] and [110] NWs using three different parameter sets. We show that although the magnitude of the mobility strongly depends on the chosen parameters, the mobility *trends* versus diameter and orientation do not change. The *relative* increase of the mobility is rather insensitive to the parameters used. As described in Refs [24, 29, 31], the most accurate deformation potentials for NWs may finally be obtained by comparing mobility with experimental data, which for NWs at this time are sparse.

The parameter sets used are:
A. $D_{ADP}$=5.39eV and $D_{ODP}$=13.24 x $10^{10}$ eV/m [24, 29]
B. $D_{ADP}$=5.0eV  and $D_{ODP}$=10.5  x $10^{10}$ eV/m [24, 31]
C. $D_{ADP}$=5.0eV  and $D_{ODP}$=9.0  x $10^{10}$ eV/m (bulk Si values) [22, 29]

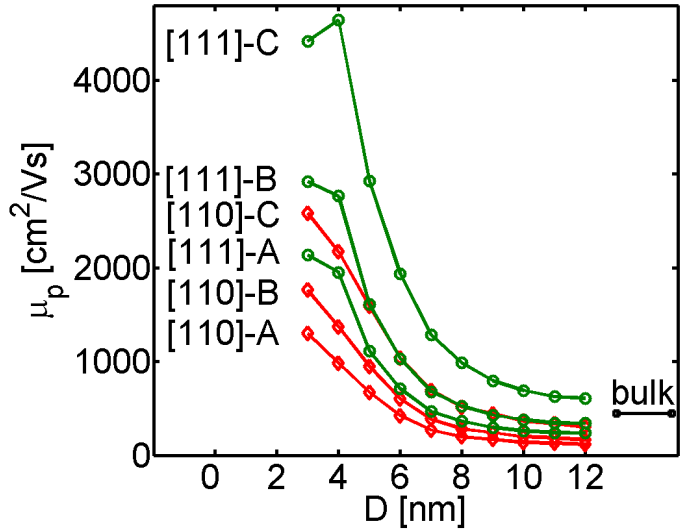

Figure A.1: The phonon limited mobility for the [110] (diamond) and [111] (circle) NWs for three different sets of deformation potential parameter sets.

Parameter set C is the values used for bulk Si. Although at larger diameters the mobility values are closer to the bulk value $\mu_p$~450cm$^2$/Vs, at smaller diameters they diverge to very large (and probably unrealistic) values. In the manuscript, the calculations are performed using set A, the most pessimistic of the three.

Figure 1:

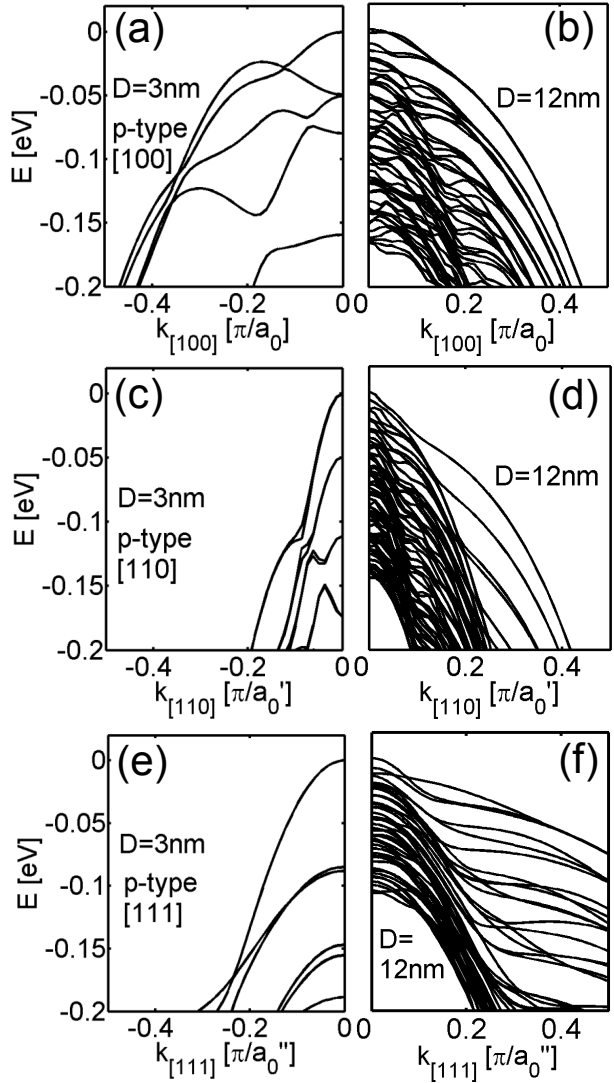

Figure 1 caption:

Dispersions of p-type NWs for various orientations and diameters (*D*). The left column is for *D*=3nm, and the right column for *D*=12nm. (a) [100], *D*=3nm. (b) [100], *D*=12nm. (c) [110], *D*=3nm. (d) [110], *D*=12nm. (e) [111], *D*=3nm. (f) [111], *D*=12nm. $a_0$, $a_0'$ and $a_0''$ are the unit cell lengths for the wires in the [100], [110], and [111] orientations, respectively.

Figure 2:

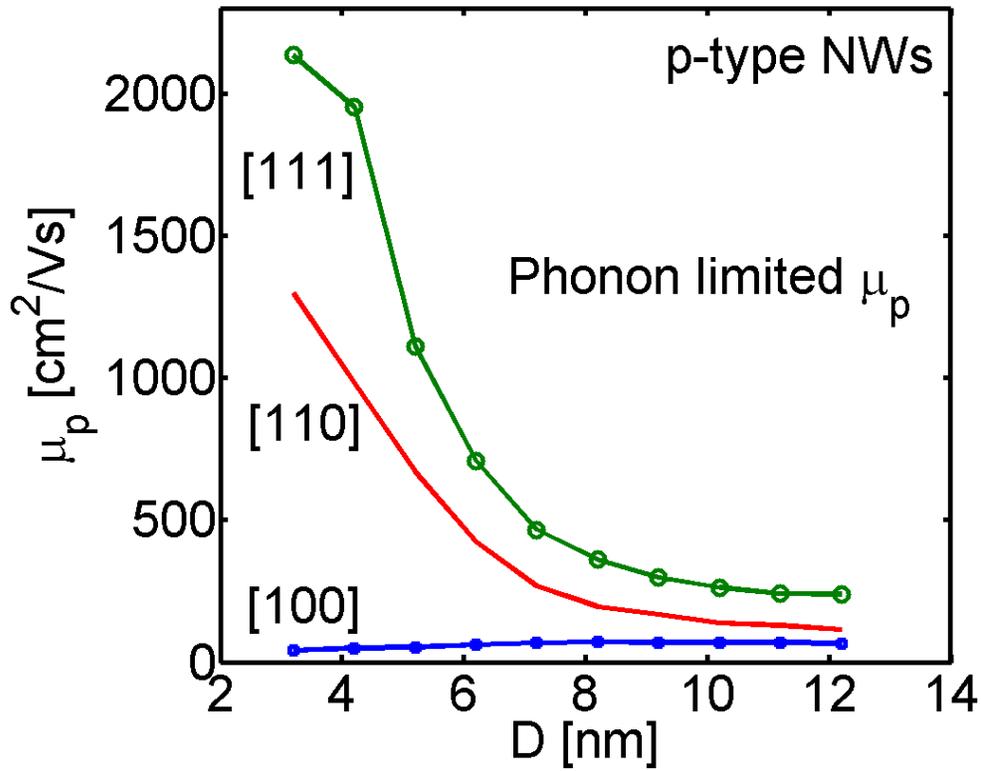

Figure 2 caption:

Hole phonon-limited mobility for cylindrical p-type NWs in [100] (dotted), [110] and [111] (circled) transport orientations as a function of the NWs' diameter. Large increases in [110] and [111] NW mobilities are observed with diameter scaling.

Figure 3:

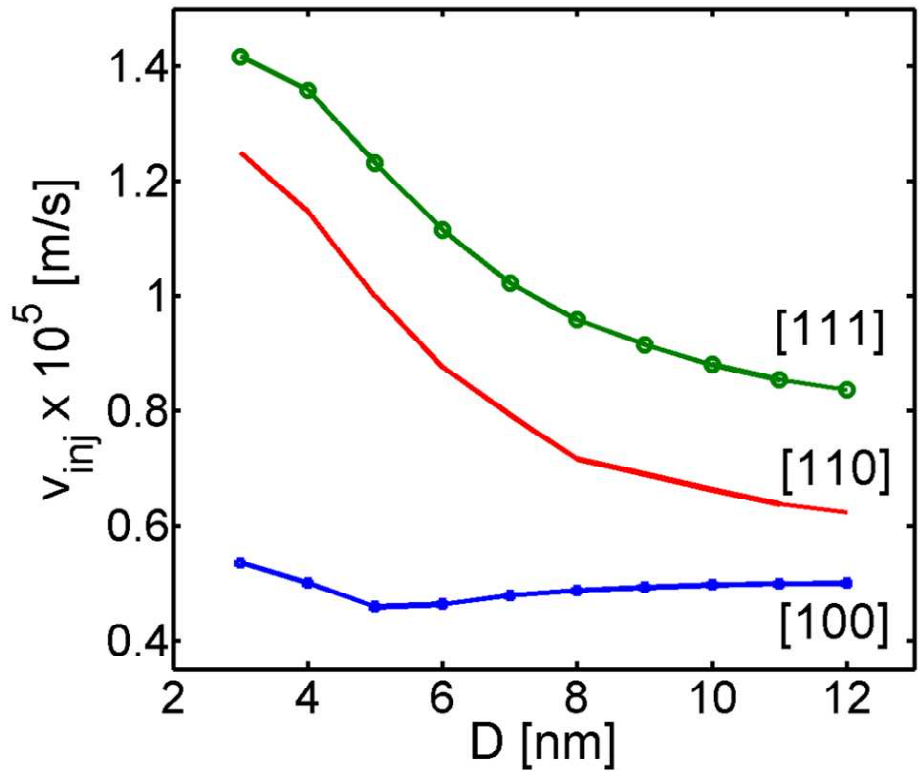

Figure 3 caption:

The carrier velocities of cylindrical p-type NWs in [100], [110], and [111] transport orientations versus diameter under non-degenerate conditions.

Figure 4:

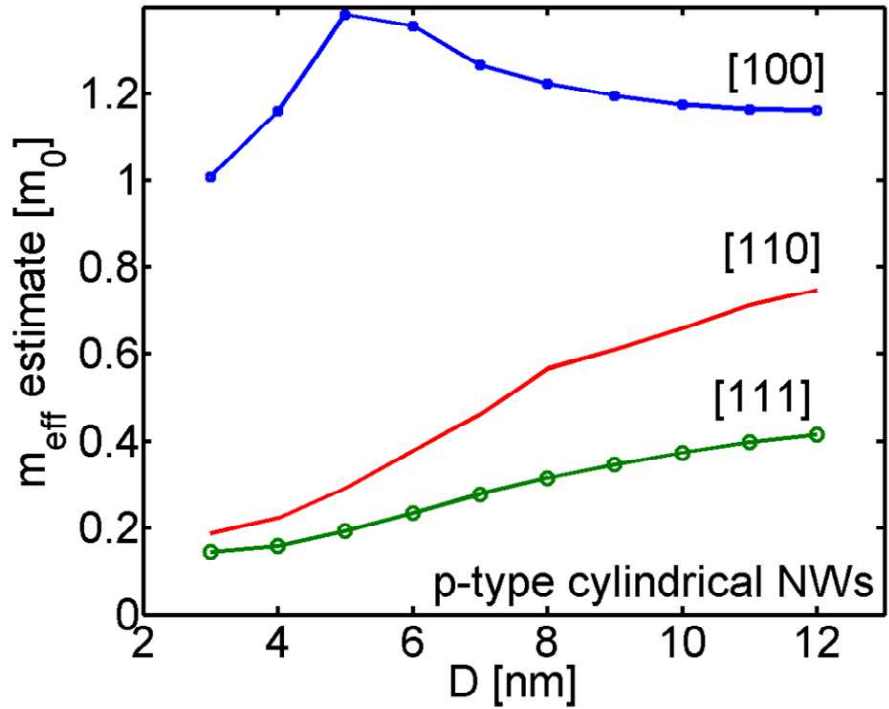

## Figure 4 caption:

An estimate of the effective mass of cylindrical p-type NWs in [100], [110], and [111] transport orientations versus diameter with all the bands in the bandstructure of the NWs considered. The results are for non-degenerate conditions.

Figure 5:

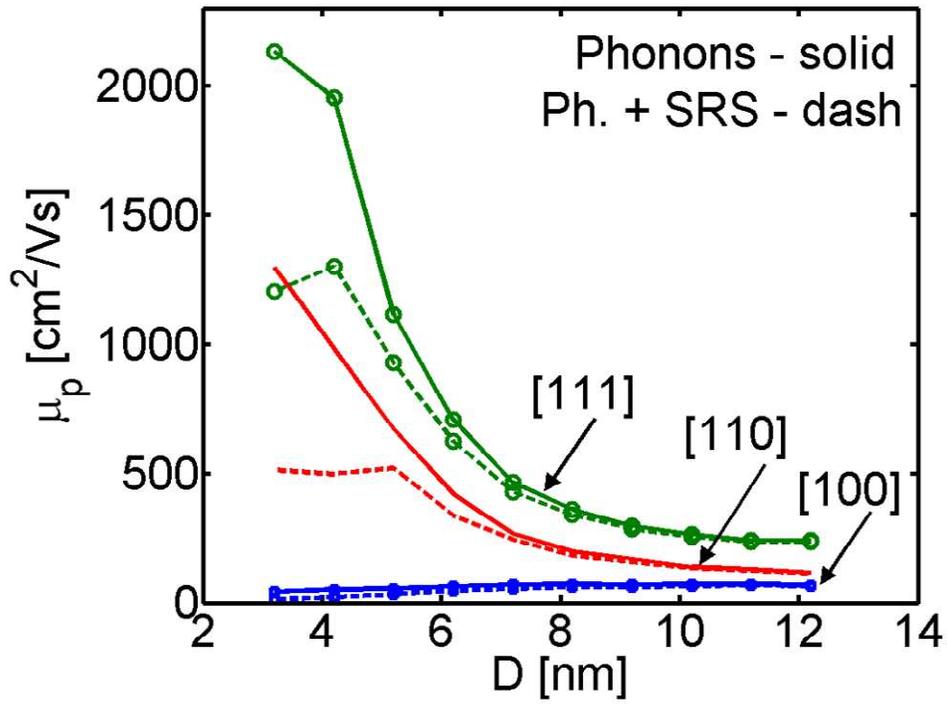

Figure 5 caption:

Hole phonon and surface roughness scattering limited mobility for cylindrical p-type NWs in [100] (dotted), [110] and [111] (circled) transport orientations as a function of the NW diameter.

Figure 6:

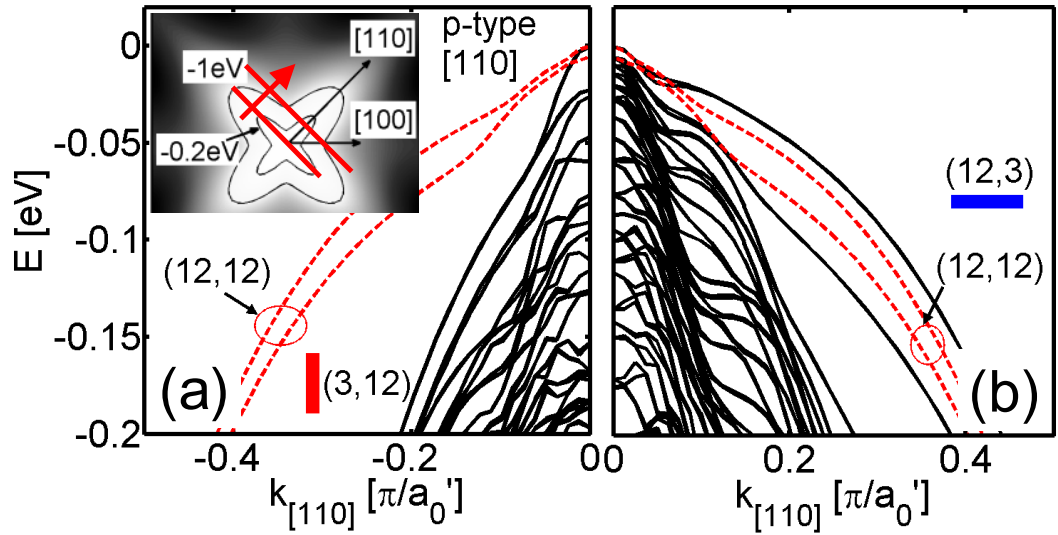

Figure 6 caption:

Dispersions of the [110] NW with different quantizations in width and height directions: (a) *W*=3nm and *H*=12nm (thin and tall). (b) *W*=12nm and *H*=3nm (thick and short). Inset of (a): The Si bulk (100) energy surface. The 45° inclined lines show the relevant energy regions for devices with physical surface confinement perpendicular the direction of the red arrow.

Figure 7:

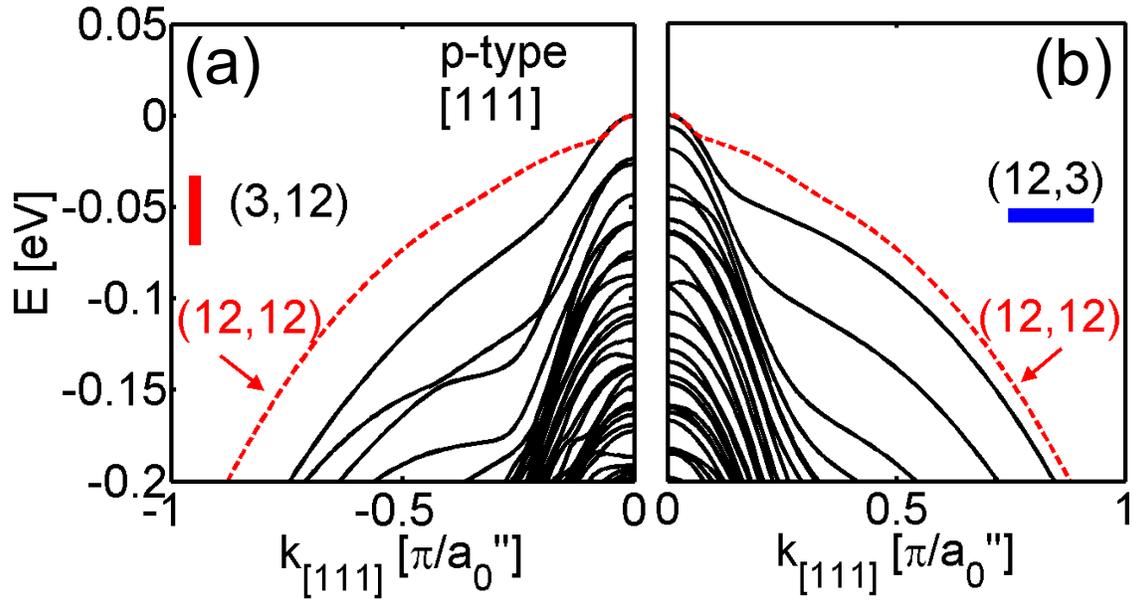

Figure 7 caption:

Dispersions of the [111] NW with different quantizations in width and height directions: (a) $W$=3nm and $H$=12nm (thin and tall). (b) $W$=12nm and $H$=3nm (thick and short). The red-dotted lines show the highest subband of the $W$=12nm, $H$=12nm square [111] NW.